# Size distribution of particles in Saturn's rings, missed moonlets and misinterpretation of Chariklo rings

Brilliantov et al. (1) propose a model for the size distribution $\sim R^{-3}$ for small particles with radius R and $\sim \exp(-(R/R_c)^3)$ for large particles, where $R_c$=5.5 m: "yet neither the power-law dependence nor the upper size cutoff have been established on theoretical ground" (1). The following comments are necessary:

In 1989 Longaretti found analytically $\sim R^{-3}$ for small particles and $R^{-6}$ for large ones (2). Similar solutions $R^{-6}$ for large bodies, and $R^{-q}$ (2.7<q<3.1) for small particles have been found in the 1990-1992 independently from (2) using different method – see (3). The law $R^{-6}$ also describes moonlets with size ~ 0.1-1 km. That allowed us to estimate the number of particles with R = 5m in A-ring as $N\sim 5*10^{13}$, $N\sim 2*10^5$ for particles with R = 125m and N~200 for R = 400m, which is comparable with observations. Cut-off law from (1) does not describe moonlets and requires new mechanism for the origin of ~ 1 km size bodies.

Collisional destruction of the large particles is ineffective if the debris stays inside the Hill's sphere of the large particle and during one revolution returns on the particle. Paper (1) does not take into account the key effect of self-gravitation of large particles. In (1) a velocity of particle does not depend on the particle size and the distance to the planet r: V=const(R, r). Consequently, the large particles with mass M and the escape velocity $V_{esc}=(GM/R)^{1/2}$ > V will grow indefinitely, contrary to the statement (1). Longaretti used a more accurate condition of destruction of particles: $\Omega R/2 > V$, where $\Omega$ – angular velocity of the rings: "A relative increase of the erosion/destruction rate of the large particles must take place, because these particles have relative velocities of collision larger than the dispersion velocity, due to the differential Keplerian motion" (2).

Decreasing $\Omega R \sim r^{-3/2}$ with distance r explains why planetary rings exist only near the planet. For the rings $\Omega R > \alpha (GM/R)^{1/2}$ where $\alpha \sim 1$: "the exterior radius $R_e$ of a planet's primary… ring system and the size of the largest member particles are determined not by tidal forces but by a more efficient mechanism: erosion of the particles surfaces through grazing collisions in the differentially rotating disk" (4). Computer simulation of the behavior of debris of destroyed particle showed that with increasing distance from the planet, most of the debris returns back to the particles (5). This effect causes a sharp transition from the zone of the rings to the area of satellites. The new model (1) does not explain the difference between the rings and the satellites and the authors (1) suggest that their calculations are applicable to the rings of the Chariklo and Chiron. In fact, the Chariklo' rings are not examples of planetary rings, but proto-satellite disks (see Figure).

Simplified physical model that does not describe moonlets and the outer boundary of the rings makes the article (1) not a step forward but a step backwards in comparison with the papers of 20-30 years old.

Nick Gorkavyi
SSAI/GSFC/NASA

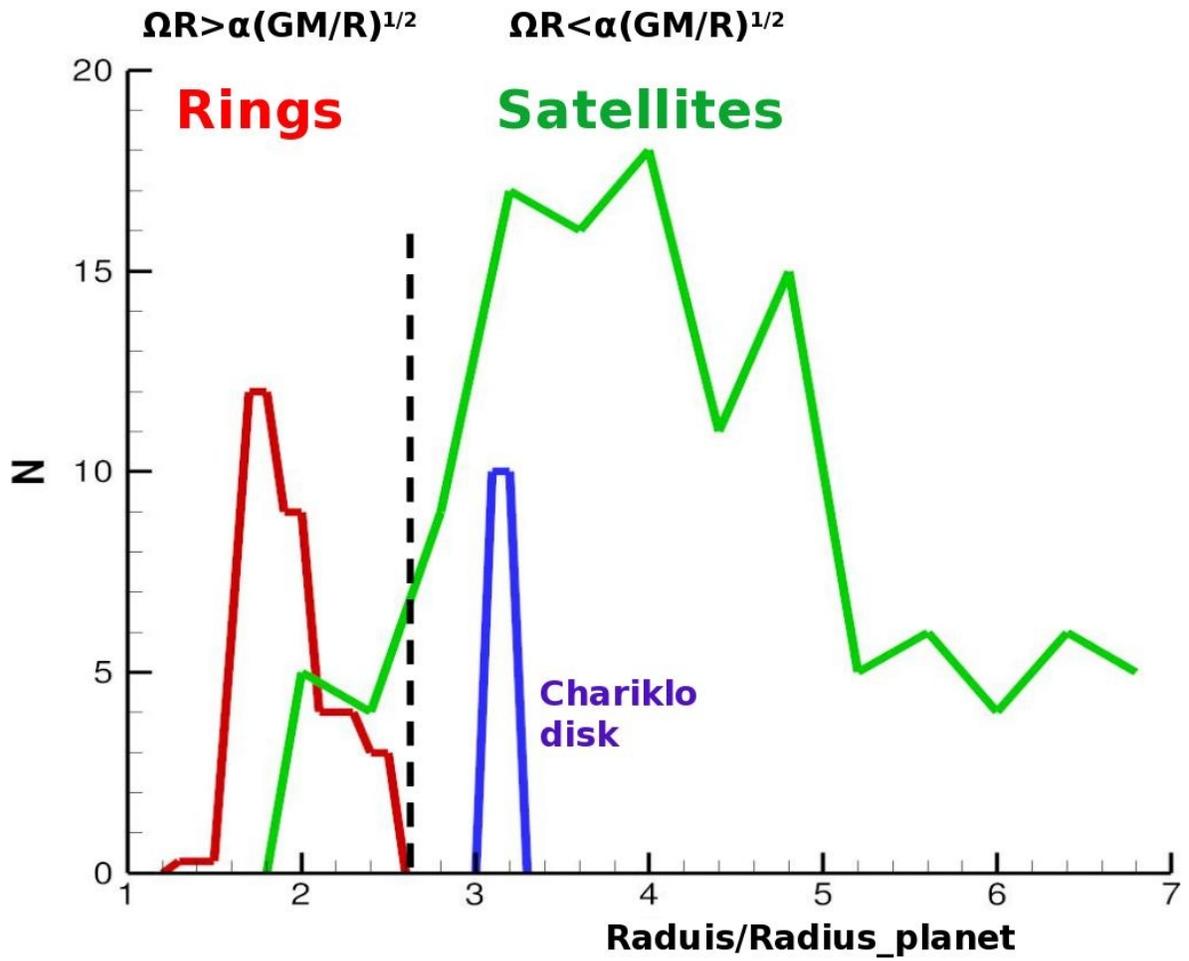

Fig. The distribution of planetary rings (red), satellites of asteroids (green) and Chariklo ring/disk (blue) by the distance from the central body.